\documentclass[preprint,preprintnumbers,amsmath,amssymb,prl]{revtex4}
\usepackage{graphicx}
\usepackage{bm}
\begin{document}
\title{Switching of a large anomalous Hall effect between metamagnetic phases of a non-collinear antiferromagnet}
\author{Christoph S\"urgers$^1$}
\email[]{christoph.suergers@kit.edu}
\author{Thomas Wolf$^2$}
\author{Peter Adelmann$^2$}
\author{Wolfram Kittler$^1$}
\author{Gerda Fischer$^1$}
\author{Hilbert v. L\"ohneysen$^{1,2}$}
\affiliation{$^1$Physikalisches Institut, Karlsruhe Institute of Technology, P.O. Box 6980, 76049 Karlsruhe, Germany}
\affiliation{$^2$Institut f\"ur Festk\"orperphysik, Karlsruhe Institute of Technology, P.O. Box 3640, 76021 Karlsruhe, Germany}

\begin{abstract}
The anomalous Hall effect (AHE), which in long-range ordered ferromagnets appears as a voltage transverse to the current and usually is proportional to the magnetization, often is believed to be of negligible size in antiferromagnets due to their low uniform magnetization. However, recent experiments and theory have demonstrated that certain antiferromagnets with a non-collinear arrangement of magnetic moments exhibit a sizeable spontaneous AHE at zero field due to a non-vanishing Berry curvature arising from the quantum mechanical phase of the electron's wave functions \cite{suzuki_large_2016,nayak_non-vanishing_2015,nakatsuji_large_2015,gomonay_berry-phase_2015,chen_anomalous_2014,shindou_orbital_2001}. Here we show that antiferromagnetic Mn$_5$Si$_3$ single crystals exibit a large AHE which is strongly anisotropic and shows multiple transitions with sign changes at different magnetic fields due to field-induced rearrangements of the magnetic structure despite only tiny variations of the total magnetization. The presence of multiple non-collinear magnetic phases offers the unique possiblity to explore the details of the AHE and the sensitivity of the Hall effect on the details of the magnetic texture.
\end{abstract}

\maketitle

Among the magnetoelectronic transport phenomena, the Hall effect takes a unique position because it provides, for single-band metals, a measure of the Fermi volume, i.e., the volume in momentum space enclosed by the Fermi surface. The Hall effect generates a voltage $V_{\rm x}$ transverse to the current $I_{\rm y}$ in a perpendicular magnetic field $H_{\rm z}$ from which the Hall resistivity $\rho_{\rm yx} =V_{\rm y} t/I_{\rm x}$ or the Hall conductivity $\sigma_{\rm xy} \approx \rho_{\rm yx}/\rho_{\rm xx}^2$ are obtained ($t$: sample thickness along $z$ direction, $\rho_{\rm xx}$: longitudinal resistivity). In ferromagnets, the AHE $\rho_{\rm yx}^{\rm AH}$, i.e., the contribution in addition to the ordinary Hall effect $\rho_{\rm yx}^{0} = R_0 B$ caused by the Lorentz force acting on the charge carriers, appears as a consequence of broken time-reversal symmetry and spin-orbit coupling (SOC) and was suggested to depend linearly on the magnetization $M$, $\rho_{\rm yx}^{\rm AH} = R_{\rm S} \mu_0M$ or $\sigma_{\rm xy}^{\rm AH} = S_{\rm H} M$ \cite{nagaosa_anomalous_2010}. However, even in materials with low or even zero magnetization, a geometrical or ``topological'' Hall effect arising solely from the electronic band structure independent of SOC is observed which is attributed to the non-vanishing Berry curvature of electrons migrating through a chiral spin texture \cite{xiao_berry_2010,machida_time-reversal_2010,schulz_emergent_2012,ueland_controllable_2012,nagaosa_topological_2013}. The Berry curvature $\bm{\Omega}_n(\bm{k})$ of the $n$th band gives rise to an anomalous velocity $\bm{v}_n({\bm k})$ of the Bloch electron in a given state $\bm k$ and a Hall current transverse to the electric field \cite{xiao_berry_2010}. 

A large AHE has been predicted to occur also in antiferromagnets with a non-collinear arrangement of magnetic moments when certain symmetries are absent and SOC is present, resulting in a Berry curvature and a sizeable AHE in zero magnetic field \cite{chen_anomalous_2014}. This has been recently proven for hexagonal Mn$_3$Ge \cite{nayak_non-vanishing_2015} and Mn$_3$Sn \cite{nakatsuji_large_2015}, showing at room temperature spontaneous Hall conductivities of 50 and 30\, $\Omega^{-1} {\rm cm}^{-1}$ in zero magnetic field, respectively. These values are of similar magnitude like in ferromagnetic metals and semiconductors. A large Hall response up to 200\, $\Omega^{-1} {\rm cm}^{-1}$ has also been reported for the half-Heusler antiferromagnet GdPtBi originating from avoided crossings or Weyl points of the electronic band structure due to the breaking of time-reversal and lattice symmetry \cite{suzuki_large_2016}. The results demonstrate that the large AHE of non-collinear antiferromagnets associated with low magnetization bears considerable potential for future applications. Thus, spintronic devices based on antiferromagnetic metals have been proposed to show current-induced phenomena like in ferromagnets and new concepts for functional devices have been developed to exploit advantages specific to antiferromagnets \cite{jungwirth_antiferromagnetic_2015,baltz_antiferromagnetism_2016,macdonald_antiferromagnetic_2011,cheng_spin_2014}.  

\begin{figure}
\centerline{\includegraphics[width=\columnwidth]{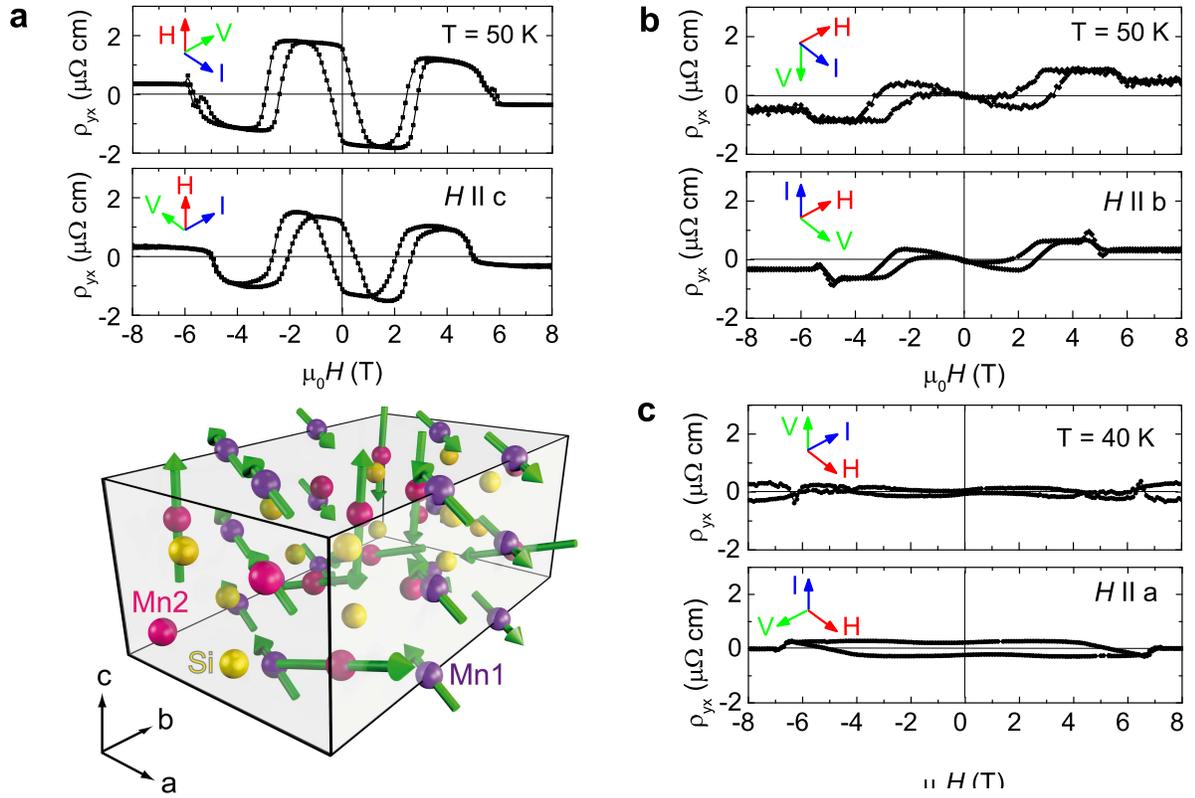}}
\caption{\textbf{$|$ Anisotropic anomalous Hall effect of Mn$_5$Si$_3$.} Schematic shows the non-collinear AF1 phase of Mn$_5$Si$_3$ below $T_{\rm N1}$ = 60 K  \cite{brown_low-temperature_1992}. Solid lines delineate the orthorhombic unit cell. Green arrows visualize the magnetic Mn moments with their relative size indicated by the arrow length. \textbf{a,b,c} Hall resistivity for the magnetic field oriented parallel to the crystallographic $c$, $b$, and $a$ axes of the orthorhombic structure, respectively.}
\label{fig1}
\end{figure}

Modifications of the AHE have been also observed for polycrystalline Mn$_5$Si$_3$ thin films in the non-collinear magnetic phase \cite{surgers_large_2014,surgers_anomalous_2016}. In contrast to Mn$_3$Ge \cite{nayak_non-vanishing_2015} and Mn$_3$Sn \cite{nakatsuji_large_2015}, Mn$_5$Si$_3$ exhibits several different antiferromagnetic phases depending on magnetic field and temperature due to competing interactions between Mn moments. Below $T_{N2}$ = 100 K, the crystal structure has orthorhombic symmetry with two inequivalent Mn sites. The magnetic structure of this AF2 phase comprises a collinear arrangement of magnetic moments Mn$_2$. Below $T_{N1} \approx $ 60 K, the magnetic structure becomes highly non-collinear by realignment of Mn$_2$ moments, splitting into Mn$_{21}$ and Mn$_{22}$ moments, and additional ordering of Mn$_1$ moments due to a magneto-structural distortion (AF1 phase, Fig. \ref{fig1}) \cite{brown_low-temperature_1992,brown_antiferromagnetism_1995,silva_magnetic_2002,gottschilch_study_2012}. The three different Mn moments each occur in a parallel and antiparallel orientation thus generating six magnetic sublattices. The non-collinearity is attributed to anisotropy and frustration. In the AF1 phase, the crystal structure can be described with orthorhombic symmetry without inversion symmetry. Both transitions have been suggested to be of first order \cite{barmin_thermal_1984,vinokurova_magnetic_1995}. 
Elastic neutron-scattering experiments have shown that the magnetic texture of the non-collinear AF1 phase strongly changes in an applied magnetic field \cite{brown_low-temperature_1992,silva_magnetic_2002}. In light of the decisive effect of magnetic texture on the Berry curvature this feature is exactly what is needed to generate large variations of the Hall effect in increasing magnetic field.

Figure \ref{fig1} shows Hall-effect measurements in the non-collinear phase of  Mn$_5$Si$_3$ single crystals with different orientations of the applied magnetic field $H$ with respect to the crystallographic axes. We observe multiple transitions with sign changes of $\rho_{\rm yx}$ and a strong anisotropy of the AHE, in particular when comparing the cases for $H \parallel c$ (Fig. \ref{fig1}a) and $H \parallel a$ (Fig. \ref{fig1}c). In addition to the jump and a hysteresis of $\rho_{\rm yx}$ around zero field, Fig. \ref{fig1}a shows two jumps of $\rho_{\rm yx}$ at higher fields in positive and negative field direction. The jump at zero field is absent for field orientations along the orthorhombic $b$ and $a$ axes (Fig. \ref{fig1}b,c). At fields above 5 T at $T$ = 50 K, the Hall effect almost vanishes for all field orientations. While the $H$ direction plays a distinctive role as just discussed, the perpendicular relative orientations of current direction and voltage drop do not entail large differences, cf. upper and lower panels in Fig. \ref{fig1}a-c.

In the following we will focus on the results for the orientation $H \parallel c$, for which a large spontaneous Hall effect at zero field, a second transition at intermediate fields, and a third transition at high fields is observed. (The behaviour of $\rho_{\rm yx}$ with the field oriented along the b axis has been reported previously \cite{surgers_anomalous_2016}, see Supplementary Information.) Figure \ref{fig2}a shows that for the field direction $\parallel c$, the Hall effect decreases with increasing temperature and vanishes above $T_{N1} \approx 60$\,K. While $\rho_{\rm yx}$ at 25 K has two remanent states at $\pm 4\, \mu\Omega {\rm cm}$ at zero field, the magnetization shows only a weak difference of $M(0) = \pm 0.03 \mu {\rm _B/f.u.}$ in the hysteresis at zero field (Fig. \ref{fig2}b, upper left inset). This difference arises from a reorientation of weakly ferromagnetically coupled Mn$_1$ moments inferred from the magnetic susceptibility of Mn$_5$Si$_3$ polycrystals, suggesting a magnetic-field induced second-order transition \cite{al-kanani_magnetic_1995}. However, from the hysteresis of $M$ and $\rho_{\rm xy}$ around zero field it appears more likely that a first-order transition occurs due to a ``switching'' of Mn$_1$ moments in a weak field which gives rise to shallow variations of $M$ but huge contributions to the AHE.  

At the second transition around 5 T at 25 K, $\rho_{\rm yx}$ switches back by the same amplitude as at zero field, whereas the corresponding change of magnetization $M(0)= \pm 0.06 \mu {\rm _B/f.u.}$ (Fig. \ref{fig2}b, lower left inset) is a factor two larger than at zero field. This transition is accompanied by a 1-\% volume compression of the crystal lattice and has been attributed to a field-induced magnetostructural modification of the AF1 phase \cite{silva_magnetic_2002,gottschilch_study_2012}. Here, the magnetic field perturbs the weak ferromagnetic coupling between neighbouring Mn$_1$ moments arranged in chains along the crystallographic $c$ axis. The order of the Mn$_1$ moments is lost but the Mn$_2$ moments are still arranged in a non-collinear fashion \cite{gottschilch_study_2012}. This magnetic phase is labeled AF1'. At 58 K, an aligned moment of $\approx 0.18\, \mu_{\rm B}$/Mn has been observed by neutron scattering \cite{silva_magnetic_2002}. In the present case, the jump observed at $\approx \pm $ 3 T and 50 K corresponds to an increase of $M$ by 0.06 $ \mu_{\rm B}$/Mn.  

The dependence of the Hall conductivity $\sigma_{xy}$ on the magnetization $M$ is shown in Fig. \ref{fig2}c, where the coloured areas indicate the different magnetic phases. Apart from the distinct hysteretic transitions, the Hall conductivity follows a linear behaviour $\sigma_{xy} = - 0.032\, {\rm V}^{-1} M$ with the same slope being observed at 25 K and 50 K, see blue broken lines in Fig. \ref{fig2}c. The linear behaviour corresponds to the roughly linearly increasing background magnetization $M(H)$ observed in the AF2 phase at $T$ = 75 K (Fig. \ref{fig2}c) and attributed to a tilting of Mn$_2$ moments in magnetic field. No hysteretic transitions have been observed above $T_{N1}$. 
The $\sigma_{\rm xy}(M)$ data at 75 K have been used to disentangle the contributions arising from the ordinary Hall effect and from the AHE \cite{surgers_large_2014}. We obtain $R_0 = 6 \times 10^{-10}\, {\rm m}^3/{\rm As}$ and $S_{\rm H} = -0.0173\, {\rm V}^{-1}$ at 75 K in agreement with values obtained for polycrystalline Mn$_5$Si$_3$ films \cite{surgers_large_2014}.   

The jump of $\rho_{\rm yx}$ back to the smooth 70-K curve occurs at $\pm 5$\, and $\pm 9$\, T for $T$ = 25 K and 50 K, respectively (Fig. \ref{fig2}a). This third transition is accompanied by a strong increase of $M$ by $\approx 2.1 (1.3) \mu_{\rm B}$/f.u. at 25 (50) K (Fig. \ref{fig2}b) due to a first-order metamagnetic transition arising from a rearrangement of Mn$_2$ moments to a magnetic state akin to the collinear AF2 phase \cite{vinokurova_magnetic_1990,al-kanani_magnetic_1995}. 
The same behaviour is observed in the reverse field direction thus leading to distinct transitions of the Hall effect at the metamagnetic transitions AF1 $\rightarrow$ AF1' and AF1' $\rightarrow$ AF2 for each magnetic field direction. At high fields, i.e., above these transitions, the $\rho_{\rm yx}(H)$ data for $T$ = 50 K and $T$ = 70 K coincide, indicating similarity of magnetic structure and/or Berry curvature. 

A strong field dependence is also observed in the magnetoresistivity $\rho_{\rm xx}(H)$, see Fig. \ref{fig2}d. $\rho_{xx}(H)$ shows minor variations at low fields but a huge jump  towards lower resistivity corresponding to a magnetoresistance ratio (MR) $\rho_{\rm xx}(H)/\rho_{\rm xx}(0)$ = 17 \%, and an almost linear decrease towards higher fields, see the data for $T$ = 50 K in Fig. \ref{fig2}d. The jump dissappears at $T_{\rm N1}$, see inset Fig. \ref{fig2}d. 

The fact that the AHE vanishes in the collinear AF2 phase at temperatures $T > T_{\rm N1}$ or at high fields suggests that all variations of $\rho_{\rm yx}$ or $\sigma_{\rm xy}$ and $M(H)$ below the high-field transition to the AF2 phase occur between two non-collinear magnetic phases AF1 and AF1' with different magnetic-moment arrangements leading to sign changes of the AHE of similar size. Hence, a full reversal of the Hall effect is observed when the magnetic structure changes between two non-collinear states by application of a magnetic field. 

The evolution of the AHE in the different magnetic states is also seen in the temperature dependence of the Hall effect at different applied magnetic fields, Fig. \ref{fig3}a. A strong variation of $\rho_{\rm yx}$ is observed for $T < T_{\rm N1}$. With decreasing temperature, $\rho_{\rm yx}(T)$ either decreases in a low field (1 T), or first increases and then decreases at intermediate field (4 T), or increases steadily for high field (8 T). The pronounced dips are artefacts due to misaligned contacts leading to a contribution from $\rho_{\rm xx}(T)$ to $\rho_{\rm yx}(T)$ which is considerably large at the metamagnetic transition, cf. Fig. \ref{fig2}d. The offset was roughly compensated in the paramagnetic state at $T$ = 110 K, see Supplementary Information, but could not be completely reduced during the temperature-dependent measurement. Fig. \ref{fig3}a (inset) shows that after zero-field cooling to 20 K and switching the magnetic field on and subsequently off in either direction generates a remanent AHE which decreases with increasing temperature to zero at $T_{\rm N2}$. The maximum of $\sigma_{\rm xy}(H=0) = 140\, \Omega^{-1}{\rm cm}^{-1}$ at $T$ = 25 K corresponds to an apparent AHE coefficient $S_{\rm H}^0=\sigma_{\rm xy}/M = - 5.1\, {\rm V}^{-1}$ in zero field, very similar to values observed for the non-collinear antiferromagnets Mn$_3$Sn ($S_{\rm H}^0= - 8.3\, {\rm V}^{-1}$) \cite{nakatsuji_large_2015} and Mn$_3$Ge ($S_{\rm H}^0= - 1.3\, {\rm V}^{-1}$) \cite{nayak_non-vanishing_2015}. $S_{\rm H}^0$ is much larger than the slope of $\sigma_{\rm xy}(M)$ arising form the background magnetization in the AF1 phase, see the blue broken lines in Fig. \ref{fig2}c. 

At temperatures below 20 K, the magnitude of the AHE decreases again. This is possibly due to a further change of the magnetic structure of Mn$_5$Si$_3$ at low temperatures which has not been investigated in detail up to now. In Mn$_5$Si$_3$ polycrystals, a small change of the magnetic susceptibility at $T$ = 30 K was tentatively interpreted as being associated with a rearrangement of weakly coupled spins in the magnetically frustrated configuration \cite{gottschilch_study_2012}. A similar temperature dependence is observed for the zero-field Hall conductivity $\sigma_{\rm xy}(H=0)$ (Fig. \ref{fig3}b) and the corresponding coercive field $H_{\rm co}$ (Fig. \ref{fig3}c). 

From the transitions observed in $\rho_{\rm yx}$, $M(H)$, and $\rho_{\rm xx}$ (see Supplementary Information), we obtain the magnetic phase diagram for Mn$_5$Si$_3$, Fig. \ref{fig4}. $T_{\rm N1}$ decreases strongly with increasing field while $T_{\rm N2}$ does not change with field. Below $T_{\rm N1}$ = 60 K, a new intermediate phase AF1' between the non-collinear phase AF1 and the collinear phase AF2 is established. This phase must also host a non-collinear magnetic structure because $\rho_{\rm yx}$ is nonzero in this regime and $\rho_{\rm yx}$ = 0 in the collinear phase. The state above $T_{\rm N1}$ or in high magnetic fields is thought to be akin to the collinear AF2 phase at $T >$\, 60 K due to the similar behaviour of $\rho_{\rm yx}(H)$, see Fig. \ref{fig2}a for $T$ = 50 K and 70 K. 


Hence, we observe multiple transitions of the Hall effect, each within a narrow field region. The data clearly demonstrate that the Berry curvature and, hence, the AHE are very sensitive to a field-induced switching of the spin-texture phase even though the magnetization shows only shallow variations due to partly compensated moments in the non-collinear antiferromagnetic phase. 

In addition to the conventional Hall effect, where current, voltage, and magnetic field are mutually oriented perpendicularly, we observe strong voltages transverse to the current when either the current, i.e., electrical field, or the voltage is oriented parallel to the magnetic field along the crystallographic $c$ axis (Figs. \ref{fig5}a and \ref{fig5}b respectively). These unusual configurations are assigned as ``longitudinal'' or ``unconventional'' Hall effects, respectively \cite{kim_dirac_2013}. The former is related to the so-called ``planar Hall effect'' which is usually maximal at an angle of 45$^{\circ}$ between the magnetic field and current direction and arises from the anisotropic magnetoresistance. The unconventional Hall effect, on the other hand, has to be attributed to effects arising from the Berry curvature. We note that in Mn$_5$Si$_3$ the unconventional Hall effect with the voltage perpendicular to the current but parallel to $H$ is only minor (Fig. \ref{fig5}c). Such unconventional Hall effects have been reported for Weyl semi-metals where anomalous magnetotransport phenomena (Adler-Bell-Jackiw anomaly) are observed due to a ``topological'' $E \cdot B$ term in the presence of weak antilocalization  \cite{kim_dirac_2013}. The existence of Weyl points or avoided crossings that develop in the electronic structure close to the Fermi level has also been suggested for the half-Heusler antiferromagnet GdPtBi exhibiting a large AHE \cite{suzuki_large_2016}. However, the observed strong unconventional Hall effect in Mn$_5$Si$_3$ could alternatively arise from the strong anisotropy of the magnetic structure and AHE, cf. Fig. \ref{fig1}. A full quantitative explanation of the size and sign of the AHE in Mn$_5$Si$_3$ with the hitherto unique feature of a sequence of noncollinear phases must await electronic band-structure calculations to obtain the Berry curvature and magnon dispersion in the different magnetic phases. This is challenging due to the magnetic superstructure in the non-collinear regime. 

Magnetic-field-induced transitions between multiple magnetic phases are typically observed in magnetically frustrated systems, where the frustration arises either from the geometry of the crystal lattice or from competing interactions between magnetic moments \cite{kawamura_universality_1998}. The example of Mn$_5$Si$_3$, where the non-collinear magnetic order is due to anisotropy and frustration, suggests that a number of similar metallic compounds with complex magnetic structures possibly exhibit large variations of the Hall response in magnetic field which makes such materials attractive for applications relying on magnetic-field induced switching of electronic transport properties.\\

\textbf{Methods}\\
The Mn$_5$Si$_3$ single crystals were obtained by a combined Bridgman and flux-growth technique using a Mn-rich self flux and a low cooling rate of 1.2$^{\circ}$C/h. The crystals were  characterized by powder x-ray diffraction, confirming the formation of the Mn$_5$Si$_3$ phase. Three cuboid pieces of mm- to sub-mm length and thickness with different orientations of the crystallographic $a_{\rm h}$ and $c_{\rm h}$ axes with respect to the sample edges were obtained after Laue diffraction. Resistivity and Hall-effect measurements were performed in a physical-property measurement system (PPMS) with the field oriented along the $z$ direction perpendicular to the sample $xy$ plane. 120-$\mu$m Cu wires were attached to the crystal with conductive silver-epoxy (EPOTEK H20E). Each sample was mounted in different orientations with respect to the magnetic field direction. Data were taken for both field directions and were symmetrized as described below. Magnetization curves were acquired in a vibrating sample magnetometer up to 12 T and in a SQUID magnetometer up to 5 T with the field applied in the same orientation as for the Hall-effect measurements.

\textbf{Acknowledgements}\\ 
We thank A. Ernst and W. D. Ratcliff for useful discussions.

\bibliography{Mn5Si3_xtals}

\begin{thebibliography}{10}
\expandafter\ifx\csname url\endcsname\relax
  \def\url#1{\texttt{#1}}\fi
\expandafter\ifx\csname urlprefix\endcsname\relax\def\urlprefix{URL }\fi
\providecommand{\bibinfo}[2]{#2}
\providecommand{\eprint}[2][]{\url{#2}}

\bibitem{suzuki_large_2016}
\bibinfo{author}{Suzuki, T.} \emph{et~al.}
\newblock \bibinfo{title}{Large anomalous {Hall} effect in a half-{Heusler}
  antiferromagnet}.
\newblock \emph{\bibinfo{journal}{Nature Phys.}}
  \textbf{\bibinfo{volume}{advance online publication}} (\bibinfo{year}{2016}).
\newblock
  \urlprefix\url{http://www.nature.com/nphys/journal/vaop/ncurrent/full/nphys3831.html}.

\bibitem{nayak_non-vanishing_2015}
\bibinfo{author}{Nayak, A.~K.} \emph{et~al.}
\newblock \bibinfo{title}{Large anomalous {Hall} effect driven by a
  nonvanishing {Berry} curvature in the noncolinear antiferromagnet
  {Mn}$_3${Ge}}.
\newblock \emph{\bibinfo{journal}{Science Advances}}
  \textbf{\bibinfo{volume}{2}}, \bibinfo{pages}{e1501870}
  (\bibinfo{year}{2016}).
\newblock \urlprefix\url{http://advances.sciencemag.org/content/2/4/e1501870}.

\bibitem{nakatsuji_large_2015}
\bibinfo{author}{Nakatsuji, S.}, \bibinfo{author}{Kiyohara, N.} \&
  \bibinfo{author}{Higo, T.}
\newblock \bibinfo{title}{Large anomalous {Hall} effect in a non-collinear
  antiferromagnet at room temperature}.
\newblock \emph{\bibinfo{journal}{Nature}} \textbf{\bibinfo{volume}{527}},
  \bibinfo{pages}{212--215} (\bibinfo{year}{2015}).
\newblock
  \urlprefix\url{http://www.nature.com/nature/journal/v527/n7577/full/nature15723.html}.

\bibitem{gomonay_berry-phase_2015}
\bibinfo{author}{Gomonay, O.}
\newblock \bibinfo{title}{Berry-phase effects and electronic dynamics in a
  noncollinear antiferromagnetic texture}.
\newblock \emph{\bibinfo{journal}{Phys. Rev. B}} \textbf{\bibinfo{volume}{91}},
  \bibinfo{pages}{144421} (\bibinfo{year}{2015}).
\newblock \urlprefix\url{http://link.aps.org/doi/10.1103/PhysRevB.91.144421}.

\bibitem{chen_anomalous_2014}
\bibinfo{author}{Chen, H.}, \bibinfo{author}{Niu, Q.} \&
  \bibinfo{author}{MacDonald, A.~H.}
\newblock \bibinfo{title}{Anomalous {Hall} {Effect} {Arising} from
  {Noncollinear} {Antiferromagnetism}}.
\newblock \emph{\bibinfo{journal}{Phys. Rev. Lett.}}
  \textbf{\bibinfo{volume}{112}}, \bibinfo{pages}{017205}
  (\bibinfo{year}{2014}).
\newblock
  \urlprefix\url{http://link.aps.org/doi/10.1103/PhysRevLett.112.017205}.

\bibitem{shindou_orbital_2001}
\bibinfo{author}{Shindou, R.} \& \bibinfo{author}{Nagaosa, N.}
\newblock \bibinfo{title}{Orbital {Ferromagnetism} and {Anomalous} {Hall}
  {Effect} in {Antiferromagnets} on the {Distorted} fcc {Lattice}}.
\newblock \emph{\bibinfo{journal}{Phys. Rev. Lett.}}
  \textbf{\bibinfo{volume}{87}}, \bibinfo{pages}{116801}
  (\bibinfo{year}{2001}).
\newblock
  \urlprefix\url{http://link.aps.org/doi/10.1103/PhysRevLett.87.116801}.

\bibitem{nagaosa_anomalous_2010}
\bibinfo{author}{Nagaosa, N.}, \bibinfo{author}{Sinova, J.},
  \bibinfo{author}{Onoda, S.}, \bibinfo{author}{MacDonald, A.~H.} \&
  \bibinfo{author}{Ong, N.~P.}
\newblock \bibinfo{title}{Anomalous {Hall} effect}.
\newblock \emph{\bibinfo{journal}{Rev. Mod. Phys.}}
  \textbf{\bibinfo{volume}{82}}, \bibinfo{pages}{1539--1592}
  (\bibinfo{year}{2010}).
\newblock \urlprefix\url{http://link.aps.org/doi/10.1103/RevModPhys.82.1539}.

\bibitem{xiao_berry_2010}
\bibinfo{author}{Xiao, D.}, \bibinfo{author}{Chang, M.-C.} \&
  \bibinfo{author}{Niu, Q.}
\newblock \bibinfo{title}{Berry phase effects on electronic properties}.
\newblock \emph{\bibinfo{journal}{Rev. Mod. Phys.}}
  \textbf{\bibinfo{volume}{82}}, \bibinfo{pages}{1959--2007}
  (\bibinfo{year}{2010}).
\newblock \urlprefix\url{http://link.aps.org/doi/10.1103/RevModPhys.82.1959}.

\bibitem{machida_time-reversal_2010}
\bibinfo{author}{Machida, Y.}, \bibinfo{author}{Nakatsuji, S.},
  \bibinfo{author}{Onoda, S.}, \bibinfo{author}{Tayama, T.} \&
  \bibinfo{author}{Sakakibara, T.}
\newblock \bibinfo{title}{Time-reversal symmetry breaking and spontaneous
  {Hall} effect without magnetic dipole order}.
\newblock \emph{\bibinfo{journal}{Nature}} \textbf{\bibinfo{volume}{463}},
  \bibinfo{pages}{210--213} (\bibinfo{year}{2010}).
\newblock
  \urlprefix\url{http://www.nature.com/nature/journal/v463/n7278/full/nature08680.html}.

\bibitem{schulz_emergent_2012}
\bibinfo{author}{Schulz, T.} \emph{et~al.}
\newblock \bibinfo{title}{Emergent electrodynamics of skyrmions in a chiral
  magnet}.
\newblock \emph{\bibinfo{journal}{Nat Phys}} \textbf{\bibinfo{volume}{8}},
  \bibinfo{pages}{301--304} (\bibinfo{year}{2012}).
\newblock
  \urlprefix\url{http://www.nature.com/nphys/journal/v8/n4/full/nphys2231.html}.

\bibitem{ueland_controllable_2012}
\bibinfo{author}{Ueland, B.~G.} \emph{et~al.}
\newblock \bibinfo{title}{Controllable chirality-induced geometrical {Hall}
  effect in a frustrated highly correlated metal}.
\newblock \emph{\bibinfo{journal}{Nature Communications}}
  \textbf{\bibinfo{volume}{3}}, \bibinfo{pages}{1067} (\bibinfo{year}{2012}).
\newblock
  \urlprefix\url{http://www.nature.com/ncomms/journal/v3/n9/full/ncomms2075.html}.

\bibitem{nagaosa_topological_2013}
\bibinfo{author}{Nagaosa, N.} \& \bibinfo{author}{Tokura, Y.}
\newblock \bibinfo{title}{Topological properties and dynamics of magnetic
  skyrmions}.
\newblock \emph{\bibinfo{journal}{Nat. Nanotechnol.}}
  \textbf{\bibinfo{volume}{8}}, \bibinfo{pages}{899--911}
  (\bibinfo{year}{2013}).
\newblock
  \urlprefix\url{http://www.nature.com/nnano/journal/v8/n12/full/nnano.2013.243.html}.

\bibitem{jungwirth_antiferromagnetic_2015}
\bibinfo{author}{Jungwirth, T.}, \bibinfo{author}{Marti, X.},
  \bibinfo{author}{Wadley, P.} \& \bibinfo{author}{Wunderlich, J.}
\newblock \bibinfo{title}{Antiferromagnetic spintronics}.
\newblock \emph{\bibinfo{journal}{Nat Nano}} \textbf{\bibinfo{volume}{11}},
  \bibinfo{pages}{231--241} (\bibinfo{year}{2016}).
\newblock
  \urlprefix\url{http://www.nature.com/nnano/journal/v11/n3/abs/nnano.2016.18.html}.

\bibitem{baltz_antiferromagnetism_2016}
\bibinfo{author}{Baltz, V.} \emph{et~al.}
\newblock \bibinfo{title}{Antiferromagnetism: the next flagship magnetic order
  for spintronics ?}
\newblock \emph{\bibinfo{journal}{arXiv:1606.04284 [cond-mat]}}
  (\bibinfo{year}{2016}).
\newblock \urlprefix\url{http://arxiv.org/abs/1606.04284}.
\newblock \bibinfo{note}{ArXiv: 1606.04284}.

\bibitem{macdonald_antiferromagnetic_2011}
\bibinfo{author}{MacDonald, A.~H.} \& \bibinfo{author}{Tsoi, M.}
\newblock \bibinfo{title}{Antiferromagnetic metal spintronics}.
\newblock \emph{\bibinfo{journal}{Philosophical Transactions of the Royal
  Society of London A: Mathematical, Physical and Engineering Sciences}}
  \textbf{\bibinfo{volume}{369}}, \bibinfo{pages}{3098--3114}
  (\bibinfo{year}{2011}).
\newblock
  \urlprefix\url{http://rsta.royalsocietypublishing.org/content/369/1948/3098}.

\bibitem{cheng_spin_2014}
\bibinfo{author}{Cheng, R.}, \bibinfo{author}{Xiao, J.}, \bibinfo{author}{Niu,
  Q.} \& \bibinfo{author}{Brataas, A.}
\newblock \bibinfo{title}{Spin {Pumping} and {Spin}-{Transfer} {Torques} in
  {Antiferromagnets}}.
\newblock \emph{\bibinfo{journal}{Phys. Rev. Lett.}}
  \textbf{\bibinfo{volume}{113}}, \bibinfo{pages}{057601}
  (\bibinfo{year}{2014}).
\newblock
  \urlprefix\url{http://link.aps.org/doi/10.1103/PhysRevLett.113.057601}.

\bibitem{surgers_large_2014}
\bibinfo{author}{S\"urgers, C.}, \bibinfo{author}{Fischer, G.},
  \bibinfo{author}{Winkel, P.} \& \bibinfo{author}{L\"ohneysen, H.~v.}
\newblock \bibinfo{title}{Large topological {Hall} effect in the non-collinear
  phase of an antiferromagnet}.
\newblock \emph{\bibinfo{journal}{Nat. Commun.}} \textbf{\bibinfo{volume}{5}}
  (\bibinfo{year}{2014}).
\newblock
  \urlprefix\url{http://www.nature.com/ncomms/2014/140305/ncomms4400/full/ncomms4400.html}.

\bibitem{surgers_anomalous_2016}
\bibinfo{author}{S\"urgers, C.}, \bibinfo{author}{Kittler, W.},
  \bibinfo{author}{Wolf, T.} \& \bibinfo{author}{L\"ohneysen, H.~v.}
\newblock \bibinfo{title}{Anomalous {Hall} effect in the noncollinear
  antiferromagnet {Mn}$_5${Si}$_3$}.
\newblock \emph{\bibinfo{journal}{AIP Advances}} \textbf{\bibinfo{volume}{6}},
  \bibinfo{pages}{055604} (\bibinfo{year}{2016}).
\newblock
  \urlprefix\url{http://scitation.aip.org/content/aip/journal/adva/6/5/10.1063/1.4943759}.

\bibitem{brown_low-temperature_1992}
\bibinfo{author}{Brown, P.~J.}, \bibinfo{author}{Forsyth, J.~B.},
  \bibinfo{author}{Nunez, V.} \& \bibinfo{author}{Tasset, F.}
\newblock \bibinfo{title}{The low-temperature antiferromagnetic structure of
  {Mn}$_5${Si}$_3$ revised in the light of neutron polarimetry}.
\newblock \emph{\bibinfo{journal}{J. Phys.: Condens. Matter}}
  \textbf{\bibinfo{volume}{4}}, \bibinfo{pages}{10025} (\bibinfo{year}{1992}).
\newblock \urlprefix\url{http://iopscience.iop.org/0953-8984/4/49/029}.

\bibitem{brown_antiferromagnetism_1995}
\bibinfo{author}{Brown, P.~J.} \& \bibinfo{author}{Forsyth, J.~B.}
\newblock \bibinfo{title}{Antiferromagnetism in {Mn}$_5${Si}$_3$: the magnetic
  structure of the {AF}2 phase at 70 {K}}.
\newblock \emph{\bibinfo{journal}{J. Phys.: Condens. Matter}}
  \textbf{\bibinfo{volume}{7}}, \bibinfo{pages}{7619} (\bibinfo{year}{1995}).
\newblock \urlprefix\url{http://iopscience.iop.org/0953-8984/7/39/004}.

\bibitem{silva_magnetic_2002}
\bibinfo{author}{Silva, M.~R.}, \bibinfo{author}{Brown, P.~J.} \&
  \bibinfo{author}{Forsyth, J.~B.}
\newblock \bibinfo{title}{Magnetic moments and magnetic site susceptibilities
  in {Mn}$_5${Si}$_3$}.
\newblock \emph{\bibinfo{journal}{J. Phys.: Condens. Matter}}
  \textbf{\bibinfo{volume}{14}}, \bibinfo{pages}{8707} (\bibinfo{year}{2002}).
\newblock \urlprefix\url{http://iopscience.iop.org/0953-8984/14/37/307}.

\bibitem{gottschilch_study_2012}
\bibinfo{author}{Gottschilch, M.} \emph{et~al.}
\newblock \bibinfo{title}{Study of the antiferromagnetism of {Mn}$_5${Si}$_3$:
  an inverse magnetocaloric effect material}.
\newblock \emph{\bibinfo{journal}{J. Mater. Chem.}}
  \textbf{\bibinfo{volume}{22}}, \bibinfo{pages}{15275--15284}
  (\bibinfo{year}{2012}).
\newblock
  \urlprefix\url{http://pubs.rsc.org/en/content/articlelanding/2012/jm/c2jm00154c}.

\bibitem{barmin_thermal_1984}
\bibinfo{author}{Barmin, S.~M.}, \bibinfo{author}{Mikhel'son, A.~V.},
  \bibinfo{author}{Sevast'yanov, A.~A.}, \bibinfo{author}{Kortov, S.~V.} \&
  \bibinfo{author}{Gel'd, P.~V.}
\newblock \bibinfo{title}{Thermal expansion of {Mn}$_5${Si}$_3$ single
  crystals}.
\newblock \emph{\bibinfo{journal}{Sov. Phys. Solid State}}
  \textbf{\bibinfo{volume}{26}}, \bibinfo{pages}{1977} (\bibinfo{year}{1984}).

\bibitem{vinokurova_magnetic_1995}
\bibinfo{author}{Vinokurova, L.}, \bibinfo{author}{Ivanov, V.} \&
  \bibinfo{author}{Kulatov, E.}
\newblock \bibinfo{title}{Magnetic phase transitions in single crystals of
  {Mn}$_5${Si}$_3$ and ({Mn}, {Fe})$_5${Si}$_3$}.
\newblock \emph{\bibinfo{journal}{Physica B: Condensed Matter}}
  \textbf{\bibinfo{volume}{211}}, \bibinfo{pages}{96--98}
  (\bibinfo{year}{1995}).
\newblock
  \urlprefix\url{http://www.sciencedirect.com/science/article/pii/092145269400953S}.

\bibitem{al-kanani_magnetic_1995}
\bibinfo{author}{Al-Kanani, H.~J.} \& \bibinfo{author}{Booth, J.~G.}
\newblock \bibinfo{title}{Magnetic field induced transitions in
  {Mn}$_5${Si}$_3$}.
\newblock \emph{\bibinfo{journal}{Journal of Magnetism and Magnetic Materials}}
  \textbf{\bibinfo{volume}{140--144}}, \bibinfo{pages}{1539--1540}
  (\bibinfo{year}{1995}).
\newblock
  \urlprefix\url{http://www.sciencedirect.com/science/article/pii/0304885394011575}.

\bibitem{vinokurova_magnetic_1990}
\bibinfo{author}{Vinokurova, L.}, \bibinfo{author}{Ivanov, V.},
  \bibinfo{author}{Kulatov, E.} \& \bibinfo{author}{Vlasov, A.}
\newblock \bibinfo{title}{Magnetic phase transitions and electronic structure
  of the manganese silicides}.
\newblock \emph{\bibinfo{journal}{Journal of Magnetism and Magnetic Materials}}
  \textbf{\bibinfo{volume}{90 \& 91}}, \bibinfo{pages}{121--125}
  (\bibinfo{year}{1990}).
\newblock
  \urlprefix\url{http://www.sciencedirect.com/science/article/pii/S030488531080040X}.

\bibitem{kim_dirac_2013}
\bibinfo{author}{Kim, H.-J.} \emph{et~al.}
\newblock \bibinfo{title}{Dirac versus {Weyl} {Fermions} in {Topological}
  {Insulators}: {Adler}-{Bell}-{Jackiw} {Anomaly} in {Transport} {Phenomena}}.
\newblock \emph{\bibinfo{journal}{Phys. Rev. Lett.}}
  \textbf{\bibinfo{volume}{111}}, \bibinfo{pages}{246603}
  (\bibinfo{year}{2013}).
\newblock
  \urlprefix\url{http://link.aps.org/doi/10.1103/PhysRevLett.111.246603}.

\bibitem{kawamura_universality_1998}
\bibinfo{author}{Kawamura, H.}
\newblock \bibinfo{title}{Universality of phase transitions of frustrated
  antiferromagnets}.
\newblock \emph{\bibinfo{journal}{J. Phys.: Condens. Matter}}
  \textbf{\bibinfo{volume}{10}}, \bibinfo{pages}{4707} (\bibinfo{year}{1998}).
\newblock \urlprefix\url{http://stacks.iop.org/0953-8984/10/i=22/a=004}.

\end{thebibliography}

\clearpage

\begin{figure}
\centerline{\includegraphics[width=\columnwidth]{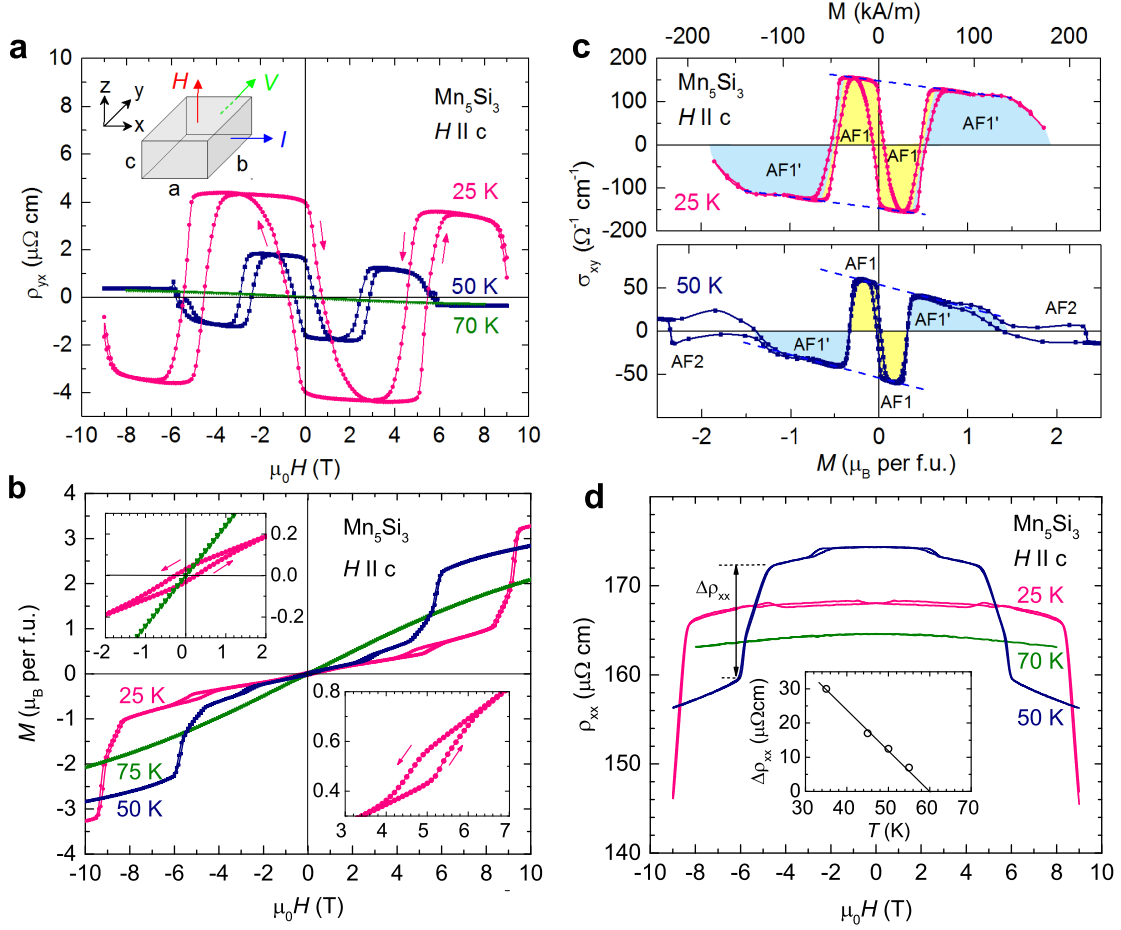}}
\caption{\textbf{$|$ Anomalous Hall effect for $H$ along $c$ and current $I$ along $a$.} \textbf{a,} Hall resistivity $\rho_{yx}(H)$ at $T$ = 25 K and 50 K (below $T_{\rm N1}$ = 60 K) and at $T$ = 70 K between $T_{\rm N1}$ and $T_{\rm N2}$. \textbf{b,} Magnetization $M$($H$) for $T$ = 25, 50, and 75 K. Insets show $M(H)$ in a small range of $H$. \textbf{c,} Hall conductivity $\sigma_{xy}$ vs. magnetization $M$. Blue broken lines indicate a linear behaviour of $\sigma_{xy}$ observed in restricted regions of $M$. Coloured areas indicate the two non-collinear magnetic phases. \textbf{d,} Magnetoresistivity $\rho_{xx}(H)$. Inset shows the temperature dependence of the jump $\Delta \rho_{xx}$ at the metamagnetic transition.}
\label{fig2}
\end{figure}

\begin{figure}
\centerline{\includegraphics[width=\columnwidth]{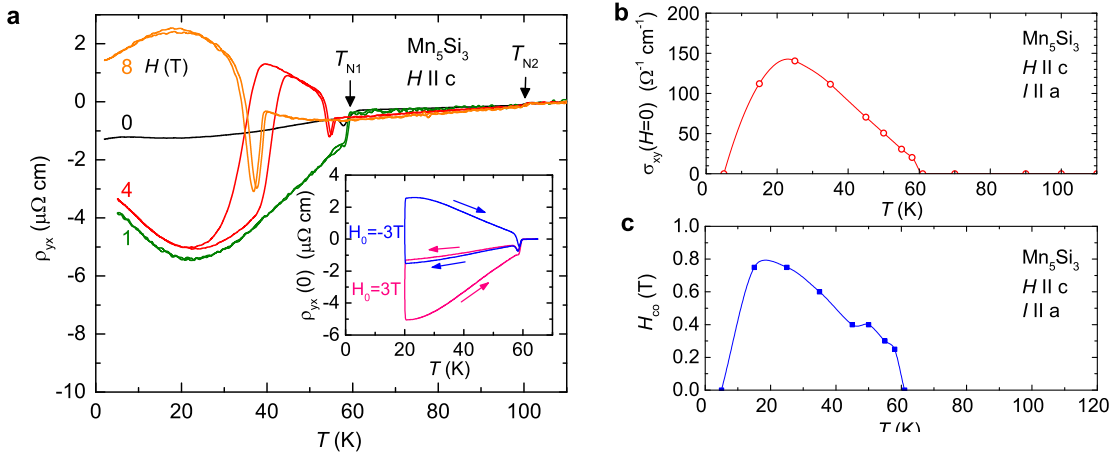}}
\caption{\textbf{$|$ Temperature dependences.} \textbf{a,} Temperature dependence of the Hall resistivity $\rho_{yx}$ in different magnetic fields. Inset shows the temperature dependence of the remanent AHE $\rho_{yx}(H=0)$ after a magnetic field of $+3$\, T or $-3$\,T was switched on and off at $T$ = 20 K. \textbf{b,} Hall conductivity at zero field $\sigma_{xy}(0) \approx \rho_{\rm yx}(0)/\rho_{\rm xx}^2$. \textbf{c,} The coercive field $H_{\rm co}$ of the Hall resistivity around zero field.}
\label{fig3}
\end{figure}

\clearpage
\begin{figure}
\centerline{\includegraphics[width=0.7\columnwidth]{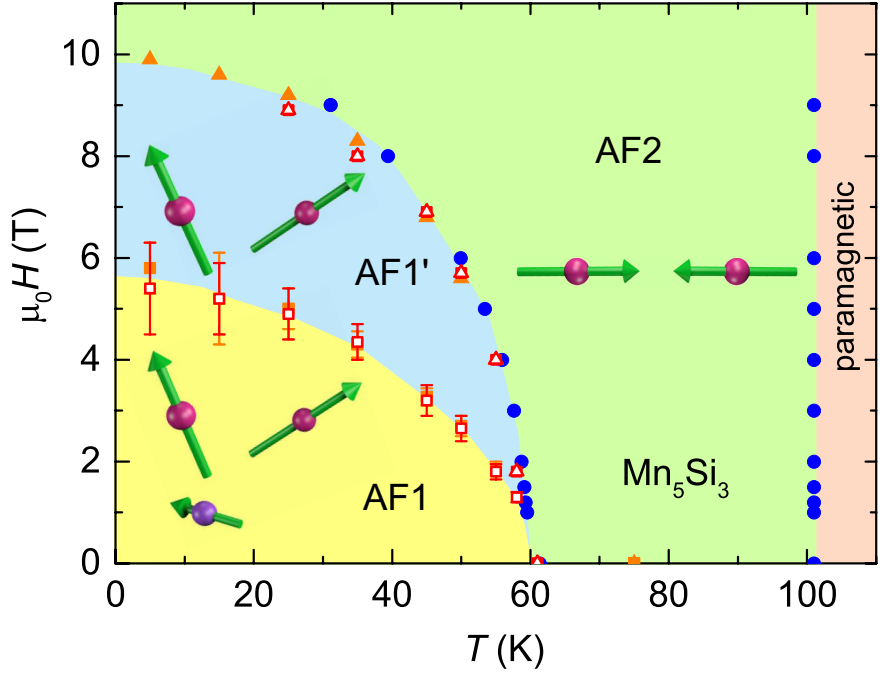}}
\caption{\textbf{$|$ Magnetic phase diagram of Mn$_5$Si$_3$.} Data obtained from measurements of Hall effect (red open symbols), resistivity (blue dots), and magnetization (orange triangles). Error bars indicate the hysteresis width of $\rho_{yx}(H)$. Arrows visualize the relative non-coplanar orientations between Mn$_1$ moments (violet) and Mn$_2$ moments (red) in the different magnetic phases. The second set of antiparallel oriented moments is not shown.}
\label{fig4}
\end{figure}

\begin{figure}
\centerline{\includegraphics[width=0.7\columnwidth]{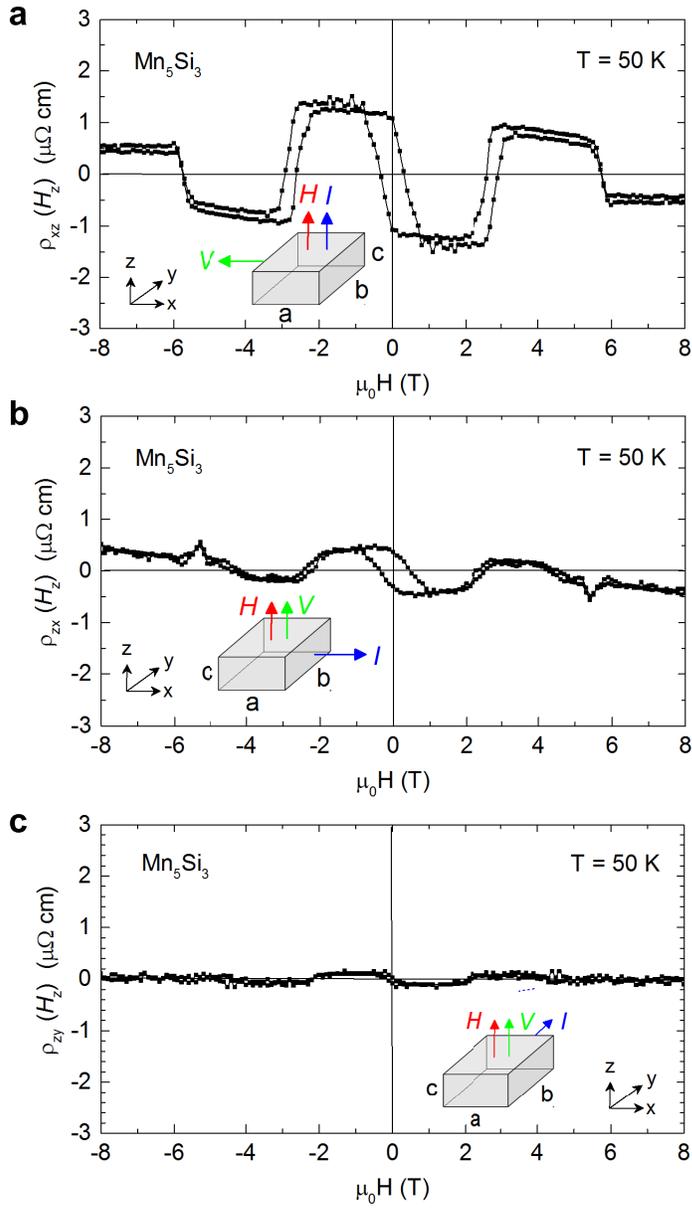}}
\caption{\textbf{$|$ Unusual Hall effects of Mn$_5$Si$_3$ below $T_{\rm N1}$ = 60 K.} Hall resistivities $\rho_{\alpha \beta}(H_{z}) = V_{\alpha}/I_{\beta}$ where either the current $I$ or the voltage $V$ transverse to the current is parallel to the magnetic field oriented along the crystallographic $c$ axis. \textbf{a,} Longitudinal Hall effect. \textbf{b,c,} Unconventional Hall effect.}
\label{fig5}
\end{figure}

\end{document}